# Spectra of the D$_2$O dimer in the O-D fundamental stretch region: the acceptor symmetric stretch fundamental and new combination bands

A. J. Barclay,[1] A. R. W. McKellar,[2] and N. Moazzen-Ahmadi[1]

[1] *Department of Physics and Astronomy, University of Calgary, 2500 University Drive North West, Calgary, Alberta T2N 1N4, Canada*

[2] *National Research Council of Canada, Ottawa, Ontario K1A 0R6, Canada*

**Abstract**

The O-D stretch fundamental region of the deuterated water dimer, (D$_2$O)$_2$, is further studied using a pulsed supersonic slit jet and a tunable optical parametric oscillator infrared source. The previously unobserved acceptor symmetric O-D stretch fundamental vibration is detected, with $K_a = 0 \leftarrow 0$ and $1 \leftarrow 0$ sub-bands at about 2669 and 2674 cm$^{-1}$, respectively. Analysis indicates that the various water dimer tunneling splittings generally decrease in the excited vibrational state, similar to the three other previously observed O-D stretch fundamentals. Two new (D$_2$O)$_2$ combination bands are observed, giving information on intermolecular vibrations in the excited O-D stretch states. The likely vibrational assignments for these and a previously observed combination band are discussed.

## I. Introduction

There have been many studies[1,2] of the water dimer using high-resolution spectroscopy during the past fifty years, and this remains an active area of experimental and theoretical[3] research. These spectra give precise and unambiguous information on intermolecular forces which provides a basis for testing theoretical calculations, and ultimately for microscopic understanding of liquid and solid water. The fundamental O-D stretch region of $(D_2O)_2$ in the mid infrared region (2600 – 2800 cm$^{-1}$) was originally investigated by Paul et al.[4,5] More recently, we published a new study[6] of this region with higher spectral resolution which enabled dimer tunneling and line width / lifetime effects to be studied in greater detail, and yielded new information on their vibrational dependence. Although there are four O-D stretch fundamental bands of $(D_2O)_2$, these previous studies only observed three of them, namely the acceptor asymmetric stretch ($\approx$2785 cm$^{-1}$), the donor free O-D stretch ($\approx$2763 cm$^{-1}$), and the donor bound O-D stretch ($\approx$2632 cm$^{-1}$).

The missing fundamental was the relatively weak acceptor symmetric stretch vibration ($\approx$2669 cm$^{-1}$), which we have finally been able to observe and analyze in the present paper. We also observe two new combination bands which give information on the intermolecular vibrations of $(D_2O)_2$. In addition to those mentioned, other studies of water dimer by high resolution mid-infrared spectroscopy include the intramolecular bending fundamentals of $(H_2O)_2$ and $(D_2O)_2$,[7,8] and the monomer O-H stretch fundamental region of $(H_2O)_2$.[9-11] In the near infrared region, there has also been work on O-H stretching overtones.[12-14]

## II. Background

The intriguing symmetry effects and tunneling dynamics of water dimer have been described many times. The reader is referred to review papers[1,2] (and the works cited therein) for more detail, and to our previous paper[6] for a simplified overview. In order to make the present paper reasonably self-contained, we present here a short recapitulation of our (already brief) previous introduction.[6]

In the water dimer equilibrium structure (Fig. 1), one monomer acts as proton (deuteron) donor and the other as acceptor. A symmetry plane contains both O atoms and the donor D atoms. These donor D atoms are inequivalent (one bound and one free) while those of the acceptor are equivalent. The structure is very floppy, with the floppiest motion, and largest tunneling splitting, corresponding to interchange of acceptor D atoms. This acceptor switching splits each rotational level into two sublevels, usually labeled "1s" (ones) and "2s" (twos). For the ground vibrational state of $(D_2O)_2$ this splitting is believed to be about 53 GHz (1.77 cm$^{-1}$),[4,15] though it is not directly measured. In this paper we assume for convenience a value of exactly 53.0 GHz.

The next largest tunneling effect involves interchange of the donor and acceptor monomers, which further splits each level into three as shown in Fig. 1. The third and smallest tunneling effect, known as bifurcation, corresponds to interchange of donor D atoms and shifts the E symmetry levels so that they are not precisely midway between their A and B partners. This shift is very small ($\approx$0.0005 cm$^{-1}$) for the ground vibrational state of $(D_2O)_2$. Water dimer vibrational modes are either in-plane ($A'$) or out-of-plane ($A''$), and in the present work we study the $A'$ acceptor symmetric O-D stretch fundamental. Acceptor switching symmetry ("1s" or "2s") flips depending on whether the vibration is $A'$ or $A''$, and also on whether $K_a$ is even or odd. Furthermore, interchange symmetry (A or B) flips depending on whether $J$ is even or odd. Here we use rotationless A/B labels which must be multiplied by the appropriate rovibrational symmetries to give the full symmetry label (see Table I of Ref. 6).

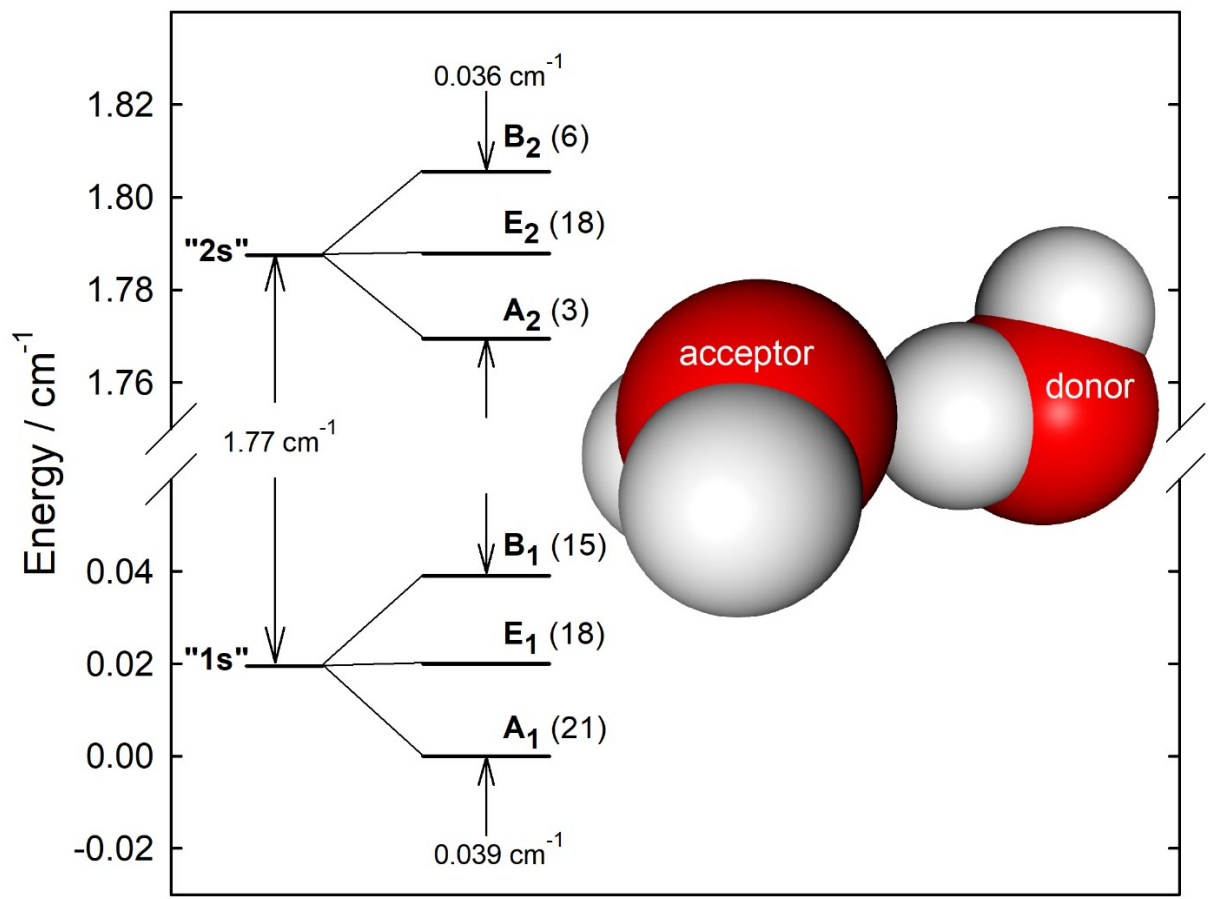

Figure 1: Ground state (J = 0, $K_a$ = 0) rotation-tunneling levels and equilibrium structure of $(D_2O)_2$. Spin statistical weights for each tunneling component are given in parentheses.

A comprehensive analysis of the $(D_2O)_2$ vibrational ground state was made by Harker et al.,[16] including new measurements and all previous spectroscopic data,[17-22] and we use their parameters here. Rotational levels are fit using the empirical expression,

$$E = \sigma + B_{av}\,[J(J+1) - K^2] - D[J(J+1) - K^2]^2 \pm [(B-C)/4][J(J+1)].$$

In this expression, $K = K_a$, and $B_{av} = (B + C)/2$. Each interchange tunneling level (A, B, E) of each acceptor switch level ("1s", "2s") of each $K_a$-value has its own origin, $\sigma$, and rotational parameters, $B_{av}$, $D$, $(B - C)$, with the $(B - C)$ term only used for $K = 1$. See Ref. 6 for more details.

As mentioned, the O atoms and donor D atoms define a plane of symmetry, and vibrations of $(D_2O)_2$ are either in-plane ($A'$ symmetry) or out-of-plane ($A''$). Since the $b$ inertial axis is perpendicular to the symmetry plane, transitions from the ground state to $A'$ vibrations can have $a$-type ($\Delta K_a = 0$) and/or $c$-type ($K_a = \pm 1$) rotational selection rules, while those to $A''$ vibrations are $b$-type ($K_a = \pm 1$).

### III. Results

Spectra were recorded as described previously,[6,23] using a pulsed supersonic slit jet expansion probed by a rapid-scan optical parametric oscillator source. The expansion mixture contained about 0.06 % $D_2O$ in helium carrier gas with a backing pressure of about 11 atmospheres (in Ref. 6, the $D_2O$ concentration was 0.07 %, not 0.007 % as mistakenly quoted). Wavenumber calibration was carried out by simultaneously recording signals from a fixed etalon and a reference gas cell. Spectral assignments and simulation were made using the PGOPHER software.[24]

### A. Acceptor symmetric stretch fundamental

As mentioned, the $(D_2O)_2$ fundamental O-D stretch region has already been studied at medium[4,5] and high[6] resolution. But previously[6] we did not detect the "missing" acceptor symmetric stretch, and it was also not observed by Paul et al.[5] (see their Fig. 1), where the authors noted the following:

"*Studies of water clusters in cryogenic matrices show that the symmetric stretch absorption intensity is substantially weaker (by about a factor of 10) than that of the antisymmetric stretch. Considering the signal-to-noise with which we observe the* [$(D_2O)_2$] *antisymmetric stretch, this reduction in intensity is sufficiently large to explain its absence in the spectrum. Nevertheless, the matrix results can be used to gain a good understanding of where the band is likely to be located in the* *gas phase. The presence of the matrix induces a remarkably uniform red-shift of 17 cm$^{-1}$ ... This locates the missing gas phase band near 2680 cm$^{-1}$."*

Part of the present spectrum is shown in Fig. 2, where we see that the parallel $a$-type ($K_a$ = 0 ← 0) component of the $(D_2O)_2$ acceptor symmetric stretch band is actually located near 2669 cm$^{-1}$, about 10 cm$^{-1}$ lower than expected. Assignment of the six tunneling components of this band was a bit confusing at first, but soon became clear. The confusion was caused by a line at 2668.912 cm$^{-1}$ (actually due to $D_2O$ monomer, and marked with an asterisk in Fig. 2) which almost seemed to form part of one series. The present assignment best explains the observed relative line intensities and also leads to reasonable values for the upper state (acceptor symmetric stretch) tunneling splittings. The spin statistical weights (Fig. 1) result in four stronger series ($A_1$, $E_1$, $B_1$, $E_2$) and two weaker ones ($A_2$, $B_2$). Two of the strong series ($A_1$ and $E_1$) show noticeable broadening due to shorter upper state lifetimes (predissociation), so their peak intensities are lower than the other strong series ($B_1$ and $E_2$). Each A and B series exhibits intensity alternation depending on even or odd values of $J''$ which helps to confirm its assignment. The results of the fit to this band are shown in Table I. The listed origin values, $\sigma$, are not band origins, but rather state origins where the ground state origins of Fig. 1 (from Harker et al.[16]) are assumed (see Table I of Ref. 6 or the present Supplementary Material for the exact values).

There were also two lines (indicated by the # symbol in Fig. 2) which lie at the expected positions of the $R(1)$ and $R(2)$ transitions of the $K = 1 \leftarrow 1$ band of the $E_2$ tunneling component. This band is relatively weak because there is little thermal population in the ground state $K = 1$ levels, and no other lines of this band were readily visible due to limited signal to noise ratio.

Another weak series of lines was assigned to $D_2O$-DOH (see upper panel of Fig. 2); this result will be reported separately together with other mixed water dimer isotopologues.

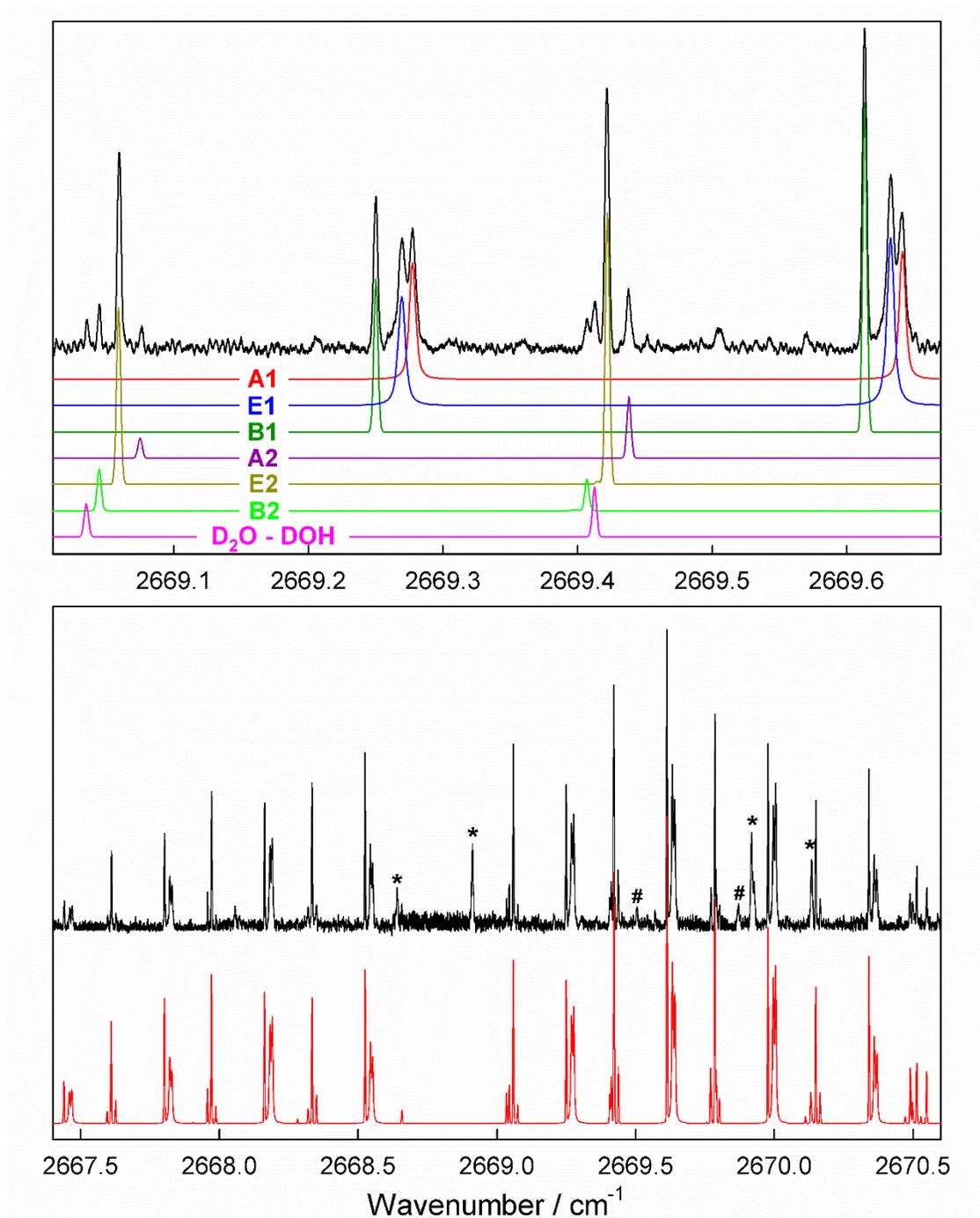

Figure 2: Observed (black) and simulated (colors) spectra of the $(D_2O)_2$ acceptor symmetric O-D stretch $K_a = 0 \leftarrow 0$ subband. The various tunneling components shown separately in the upper panel simulation and their sum is shown in the lower panel. Asterisks mark $D_2O$ monomer lines. The symbol # marks two weak lines assigned to the $K_a = 1 \leftarrow 1$ subband (see text).

Table I. Upper state parameters for the acceptor symmetric stretch fundamental band of $(D_2O)_2$ (in cm$^{-1}$). [a]

| $K_a$ | Sym | Origin, σ | $B_{av}$ | Width |
|---|---|---|---|---|
| 0 | $A_1$ | 2668.9146(3) | 0.181477(25) | 0.0032 |
| 0 | $E_1$ | 2668.9273(3) | 0.181385(30) | 0.0041 |
| 0 | $B_1$ | 2668.9270(2) | 0.181370(12) | <0.001 |
| 0 | $A_2$ | 2670.4825(3) | 0.181361(31) | <0.001 |
| 0 | $E_2$ | 2670.4847(2) | 0.181376(13) | <0.001 |
| 0 | $B_2$ | 2670.4880(4) | 0.181343(54) | <0.001 |
| 1 | $A_1$ | 2674.10 | 0.18131 | 0.045 |
| 1 | $E_1$ | 2674.102 | 0.18131 | 0.012 |
| 1 | $B_1$ | 2674.098 | 0.18131 | 0.012 |
| 1 | $A_2$ | 2673.5318 | 0.18125 | 0.0025 |
| 1 | $E_2$ | 2673.5422(3) | 0.181234(20) | 0.0020 |
| 1 | $B_2$ | 2673.5412(4) | 0.181286(24) | <0.001 |

[a] Uncertainties in parentheses are 1σ in units of the last quoted decimal. All centrifugal distortion parameters $D$ were fixed at an average ground state value of $1.2 \times 10^{-6}$ cm$^{-1}$. Asymmetry doubling parameters $(B-C)/4$ were fixed at average ground state values of 5.5 or $9.9 \times 10^{-4}$ cm$^{-1}$ for the 1s or 2s states, respectively. Parameters for the broad $A_1$, $E_1$, and $B_1$ states with $K = 1$ were hand-fitted and are only approximate. Parameters for the $A_2$ state with $K = 1$ were hand-fitted because it is weak and only barely resolved from the $E_2$ state with $K = 1$. Width values for $K_a = 1$ apply to the $P$- and $R$-branches.

The $(D_2O)_2$ acceptor symmetric stretch perpendicular $c$-type bands with $K_a = 1 \leftarrow 0$ are shown in Fig. 3. Transitions in the 1s tunneling band are significantly broadened, especially the $A_1$ component, while the 2s band is sharper but still has noticeable broadening for the $A_2$ and $E_2$ components. The results of our fit of these bands are given in the lower part of Table I. The vibrational origin for the $A_1$ $K = 1$ upper state and the $B'$-values for all the 1s states are not well determined because of the large width of these transitions. As well, the $B'$-value for $A_2$ is not well determined because its transitions were only barely resolved from the stronger $E_2$ transitions.

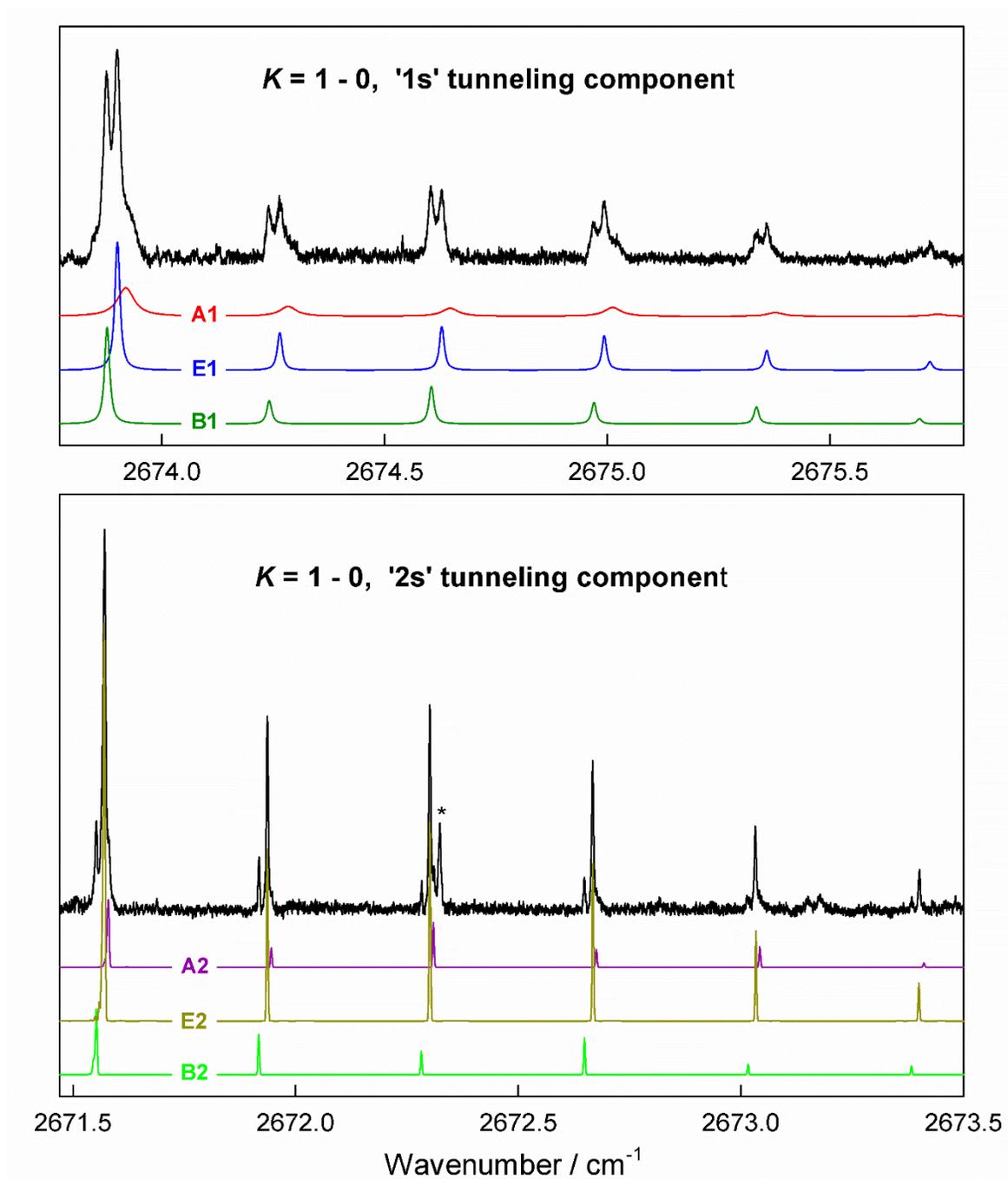

Figure 3: Observed (black) and simulated (colors) spectra of the Q- and R-branch regions of the $(D_2O)_2$ acceptor symmetric O-D stretch sub-bands with $K_a = 1 \leftarrow 0$.

### B. Combination Bands

The intermolecular vibrational modes of $(D_2O)_2$ have been explored by means of high resolution far infrared spectroscopy[25,26] and *ab initio* theory.[15,27] There are 6 such modes which may be described as (in approximate order of increasing frequency): donor torsion ($A''$ symmetry), acceptor wag ($A'$), acceptor twist ($A''$), intermolecular stretch ($A'$), in-plane bend ($A'$), and out-of-plane bend ($A''$). Combining these six intermolecular modes with the four O-D stretch fundamentals leads to a large number of possible combination bands in the present spectral region!

We previously observed[6] a very weak combination band of $(D_2O)_2$ at 2753.9 cm$^{-1}$ and assigned it as a 1s $K_a = 1 \leftarrow 0$ transition. At that time, the most likely vibrational assignment seemed to be the combination of the donor bound O-D stretch fundamental plus the donor twist overtone intermolecular mode, though this was not certain. However, now we know that the true value of the acceptor symmetric stretch fundamental is lower than expected, and a more likely assignment can be made to this symmetric stretch plus the acceptor wag intermolecular mode. The $K_a = 1$ level of the wag in the excited symmetric stretch state would then have an origin of 85.2 cm$^{-1}$, very close to its ground state value[15] of 85.57 cm$^{-1}$. The difference between the two of course includes both the change in acceptor wag frequency, the change in acceptor switch splitting, and the change in $K_a = 0$ to 1 interval (i.e. the $A$ value) between the ground and excited states. Our analysis of this band[6] seemed to indicate that it was *c*-type (based on the shape of the partially resolved $Q$-branch), which could argue against the assignment just given since the combination of symmetric stretch plus acceptor wag has $A''$ symmetry. In fact, however, this distinction between *b*- and *c*-type was not at all well established.[6]

A new combination band observed near 2844.5 cm$^{-1}$ is illustrated in Fig. 4. We assigned it as a 1s $K_a = 0 \leftarrow 0$ transition and analyzed it with results as shown in Table II and in the Fig. 4 simulation. The vibrational assignment for this band is almost certainly the donor free O-D stretch plus the acceptor wag intermolecular mode. The combination band origin minus the donor free O-D stretch origin gives us a value of 81.09 cm$^{-1}$ for the excited state acceptor wag frequency, in good agreement with the ground state value[15] of 82.64 cm$^{-1}$. The results of our fit to this band are listed in Table II.

Another combination band was observed near 2872.8 cm$^{-1}$, as shown in the lower panel of Fig. 4. Only two series of lines were observed and we think that their most likely assignments are to the E$_1$ and B$_1$ tunneling states as given in Table II. Perhaps the most likely explanation for the apparent absence of the A$_1$ series is that its upper state levels are highly broadened by predissociation. There are two possible vibrational assignments for this band, either as the donor free O-D stretch plus the donor twist overtone[15] (called DT2, and earlier[25] thought to be the in-plane bend), or else as the acceptor asymmetric O-D stretch plus the acceptor twist intermolecular mode.[15] If the excited state intermolecular frequencies were the same as in the ground state, then these two possibilities would be predicted at 2867.7 and 2877.7 cm$^{-1}$, respectively. How to choose between them? We note (Table II) that $B$ values are significantly increased in the upper state, and this favors the latter assignment since the acceptor twist is known to have a large $B$ value. More specifically, assuming simple additivity for the $\alpha$'s (changes in $B$ values) of the various vibrational modes, we predict $B \approx 0.1818$ or 0.1835 cm$^{-1}$, respectively, for the two assignments, based on ground state values for the intermolecular modes.[25,26] The observed $B$ values for the two series of the 2872.8 cm$^{-1}$ band are 0.1834 and 0.1837 cm$^{-1}$, clearly supporting the latter assignment (asymmetric stretch plus acceptor twist).

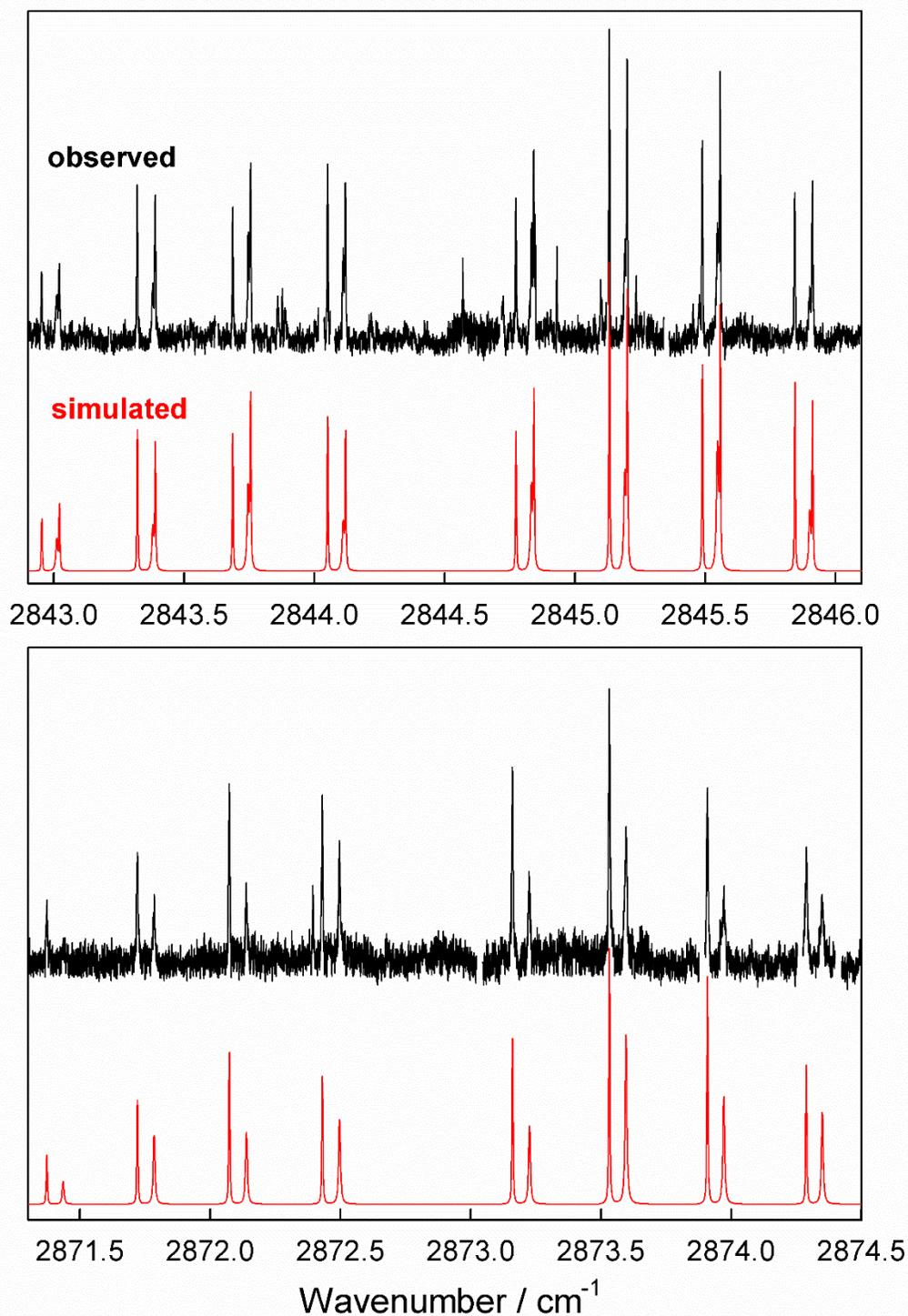

Figure 4: Observed and simulated spectra of two (D₂O)₂ combination bands. We assign the 2844.5 cm⁻¹ band (upper panel) as the donor free O-D stretch intramolecular mode plus acceptor wag intermolecular mode. We assign the 2872.8 cm⁻¹ band (lower panel) as the acceptor asymmetric stretch intramolecular mode plus acceptor twist intermolecular mode.

Table II. Upper state parameters for two combination bands of $(D_2O)_2$ (in $cm^{-1}$). [a]

| $K_a$ | Sym | Origin, σ | $B_{av}$ | $D$ | Width |
|---|---|---|---|---|---|
| 0 | $A_1$ | 2844.4730(2) | 0.180245(47) | $4.2(23)\times10^{-6}$ | 0.008 |
| 0 | $E_1$ | 2844.5015(2) | 0.180193(29) | $-3.7(48)\times10^{-6}$ | 0.002 |
| 0 | $B_1$ | 2844.4521(2) | 0.180264(20) | $-2.5(10)\times10^{-6}$ | 0.002 |
| 0 | $E_1$ | 2872.8128(2) | 0.183690(37) | $9.5(13)\times10^{-6}$ | 0.007 |
| 0 | $B_1$ | 2872.8978(2) | 0.183361(68) | $5.2(33)\times10^{-6}$ | 0.001 |

[a] Uncertainties in parentheses are 1σ in units of the last quoted decimal.

This means that the acceptor twist frequency in the upper state is 88.07 $cm^{-1}$ (average of $E_1$ and $B_1$), as compared to 92.91 $cm^{-1}$ in the ground state,[15] a relatively large change which could in part be related to a change in acceptor switch splitting. The ground state twist frequency for the 2s state is 90.37 $cm^{-1}$, so the change would be smaller if our assignment of the 2872.8 $cm^{-1}$ band was changed to 2s instead of 1s. But it is difficult to simulate the observed spectrum convincingly as a 2s transition.

Incidentally, the test which assumes additivity of alpha values also works well for the 2844.5 $cm^{-1}$ band, where the vibrational assignment already seems obvious. However, in the case of the 2753.9 $cm^{-1}$ band, the test is not very useful because the upper state $B$-values are not well determined experimentally and those for the donor twist overtone in the ground state are not known for $K_a = 1$.

### IV. Discussion and conclusions

### A. Tunneling splittings

The upper state tunneling splittings given for the acceptor symmetric stretch fundamental in Table I are shown together with those of the ground vibrational state in Fig. 5 (the ground state "1s" to "2s" splitting is assumed to be exactly 53.0 GHz for $K = 0$). This figure may be compared with the results for the other O-D stretch fundamentals in Figs. 7 and 8 of Ref. 6. All excited state tunneling splittings are reduced compared to the ground state, as previously observed for the other O-D stretch fundamentals.[6] This is particularly true for the donor-acceptor interchange (A – E – B) splittings, as expected since interchange becomes more difficult in the excited state because the vibrational excitation (O-D stretch) has to be transferred as part of the interchange.[8,9,28,29] Generally speaking, the excitation of O-D stretch vibrations has significant effects on both the interchange and bifurcation tunneling motions in $(D_2O)_2$.[6]

### B. Line widths and vibrational shifts

The line width estimates in Tables I and II come from fits which assume a Gaussian instrumental width profile (0.0028 cm$^{-1}$), arising mainly from Doppler broadening due to non-orthogonality of the laser beam and jet propagation as well as smaller laser jitter/drift effects. Additional Lorentzian broadening is then included for each symmetry and $K_a$ value to account for upper state lifetime (predissociation) broadening. These Lorentzian widths are only approximate (especially in Table II) because we do not know the precise instrumental contribution and because there may also be $J$-dependence, but they are still meaningful, especially relative to each other.

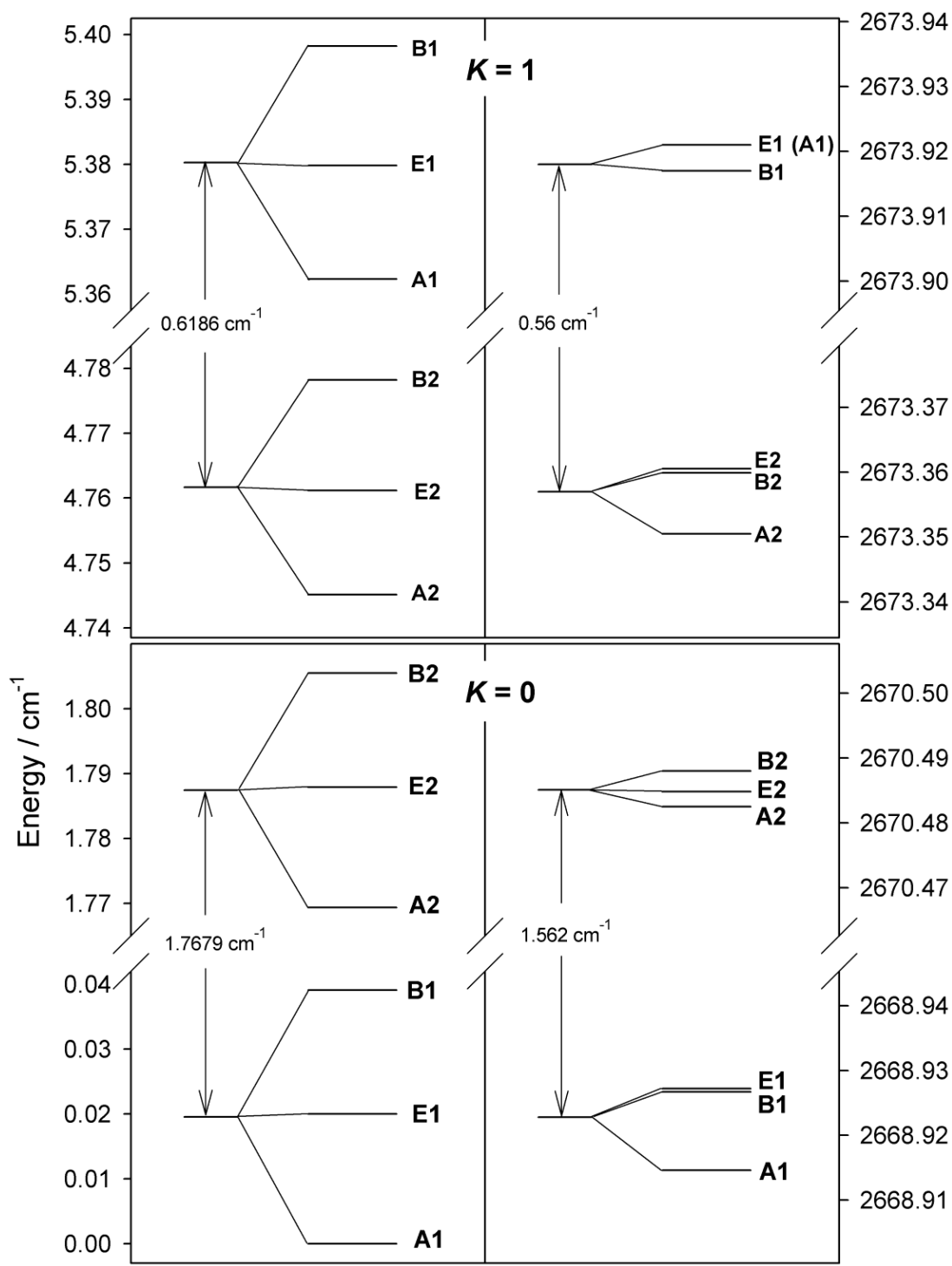

Figure 5: Comparison of (D₂O)₂ tunneling splittings in the ground vibrational state (left) and the excited acceptor symmetric O-D stretch state (right). Note the reduced donor-acceptor interchange splittings (A, E, B) in the excited state. The precise excited state $K_a$ = 1, $A_1$ level position is uncertain due to its broad observed width ($\approx$0.04 cm⁻¹).

For the symmetric stretch upper state with $K = 0$, the $B_1$, $A_2$, $E_2$, and $B_2$ series are all sharp, showing no detectible Lorentzian broadening, while the $A_1$ and $E_1$ series show broadenings of about 0.003 and 0.004 cm$^{-1}$, respectively. These effects are easily visible in the spectrum itself (Fig. 2). The upper state with $K = 1$ shows large broadening for $E_1$, $B_1$ (0.01 cm$^{-1}$) and especially $A_1$ (0.04 cm$^{-1}$), and smaller broadening for the 2s series.

The $(D_2O)_2$ acceptor symmetric stretch frequency, spin weight averaged over the six tunneling states, is about 2668.84 cm$^{-1}$, which represents a red shift of only about -2.81 cm$^{-1}$ relative to the free $D_2O$ monomer.[30] The average $K = 1 – 0$ intervals in the excited state are about 5.00 and 2.97 cm$^{-1}$ for the 1s and 2s manifolds, respectively, which represent changes of about -0.36 and -0.10 cm$^{-1}$ with respect to the ground vibrational state.

## C.  Conclusions

The "missing" acceptor symmetric stretch fundamental band of $(D_2O)_2$ has been observed at about 2669 cm$^{-1}$, a value which is somewhat lower than expected[5] but shifted by only -2.8 cm$^{-1}$ with respect to $D_2O$ itself. The predominant acceptor switching and donor-acceptor interchange tunneling splittings were found to decrease upon excitation of the symmetric O-D stretch, and a range of predissociation broadening was observed, just as noted previously[6] for the other O-D stretch fundamentals. Two new $(D_2O)_2$ combination bands were observed, adding to one such previous result. With their most likely assignments, these bands give intermolecular vibrational frequencies in the excited O-D stretch states which are very similar to their ground state values in two cases (<2 %), and somewhat different in the third case (5.5 %). We note that high resolution data comparable to that presented here and in Ref. 6 are not available for $(H_2O)_2$ itself because of large broadening effects (short lifetimes) in most of its O-H stretch fundamentals.[9,10] Fortunately, however, we have been able to observe O-D stretch spectra due to most of the ten possible mixed (H/D) water dimer isotopologues and this will be the subject of a future publication.

**Supplementary Material**

Supplementary Material includes tables giving assumed ground state energies and fitted excited state energies for the acceptor symmetric stretch.

**Acknowledgements**

The financial support of the Natural Sciences and Engineering Research Council of Canada is gratefully acknowledged.

**AUTHOR DECLARATIONS**

**Conflict of Interest**

The authors have no conflicts to disclose.

Supplementary Data for:

Spectra of the $D_2O$ dimer in the O-D fundamental stretch region:

the acceptor symmetric stretch fundamental and new combination bands.

By: A.J. Barclay,[1] A.R.W. McKellar,[2] and N. Moazzen-Ahmadi[1]

[1] *Department of Physics and Astronomy, University of Calgary, 2500 University Drive North West, Calgary, Alberta T2N 1N4, Canada*

[2] *National Research Council of Canada, Ottawa, Ontario K1A 0R6, Canada*

Table A-1. Symmetry labels and assumed ground state origins for $(D_2O)_2$ (in $cm^{-1}$).

| Present notation | Complete label | $K = 0$ origin | $K = 1$ origin |
|---|---|---|---|
| $A_1$ | $A_1^+/B_1^-$ | 0.00000 | 5.36232 |
| $E_1$ | $E_1^+/E_1^-$ | 0.02002 | 5.37980 |
| $B_1$ | $B_1^+/A_1^-$ | 0.03910 | 5.39822 |
| $A_2$ | $A_2^-/B_2^+$ | 1.76937 | 4.74511 |
| $E_2$ | $E_2^-/E_2^+$ | 1.78791 | 4.76119 |
| $B_2$ | $B_2^-/A_2^+$ | 1.80551 | 4.77821 |

Table A-2. Assumed ground state energies for $K = 0$ levels of $(D_2O)_2$ (in $cm^{-1}$) (based on Table 3a of Harker et al., Mol. Phys. **105**, 497 (2007)).

| $J$ | A1 | E1 | B1 | A2 | E2 | B2 |
|---|---|---|---|---|---|---|
| 0 | 0.0000 | 0.0200 | 0.0391 | 1.7694 | 1.7879 | 1.8055 |
| 1 | 0.3624 | 0.3824 | 0.4015 | 2.1318 | 2.1503 | 2.1679 |
| 2 | 1.0872 | 1.1072 | 1.1262 | 2.8566 | 2.8751 | 2.8927 |
| 3 | 2.1744 | 2.1943 | 2.2133 | 3.9437 | 3.9622 | 3.9797 |
| 4 | 3.6237 | 3.6436 | 3.6626 | 5.3931 | 5.4115 | 5.4290 |
| 5 | 5.4353 | 5.4550 | 5.4739 | 7.2046 | 7.2230 | 7.2404 |
| 6 | 7.6087 | 7.6284 | 7.6472 | 9.3781 | 9.3964 | 9.4138 |
| 7 | 10.1440 | 10.1636 | 10.1823 | 11.9134 | 11.9316 | 11.9490 |

Table A-3:  Assumed ground state energies for $K = 1-$ levels of $(D_2O)_2$ (in cm$^{-1}$) (based on Table 3a of Harker et al., Mol. Phys. **105**, 497 (2007)).

| $J$ | A1 | E1 | B1 | A2 | E2 | B2 |
|-----|--------|--------|--------|--------|--------|--------|
| 1 | 5.5430 | 5.5605 | 5.5789 | 4.9253 | 4.9413 | 4.9584 |
| 2 | 6.2668 | 6.2842 | 6.3026 | 5.6478 | 5.6639 | 5.6809 |
| 3 | 7.3524 | 7.3698 | 7.3881 | 6.7316 | 6.7476 | 6.7646 |
| 4 | 8.7997 | 8.8171 | 8.8354 | 8.1765 | 8.1925 | 8.2095 |
| 5 | 10.6086 | 10.6259 | 10.6442 | 9.9823 | 9.9983 | 10.0153 |
| 6 | 12.7790 | 12.7963 | 12.8145 | 12.1491 | 12.1651 | 12.1820 |
| 7 | 15.3107 | 15.3279 | 15.3460 | 14.6766 | 14.6925 | 14.7094 |

Table A-4:  Assumed ground state energies for $K = 1+$ levels of $(D_2O)_2$ (in cm$^{-1}$) (based on Table 3a of Harker et al., Mol. Phys. **105**, 497 (2007)).

| $J$ | A1 | E1 | B1 | A2 | E2 | B2 |
|-----|--------|--------|--------|--------|--------|--------|
| 1 | 5.5441 | 5.5616 | 5.5800 | 4.9272 | 4.9433 | 4.9603 |
| 2 | 6.2701 | 6.2875 | 6.3059 | 5.6538 | 5.6698 | 5.6868 |
| 3 | 7.3590 | 7.3764 | 7.3947 | 6.7435 | 6.7595 | 6.7765 |
| 4 | 8.8107 | 8.8281 | 8.8464 | 8.1963 | 8.2124 | 8.2292 |
| 5 | 10.6252 | 10.6425 | 10.6607 | 10.0122 | 10.0281 | 10.0449 |
| 6 | 12.8022 | 12.8195 | 12.8376 | 12.1909 | 12.2068 | 12.2235 |
| 7 | 15.3416 | 15.3588 | 15.3769 | 14.7323 | 14.7481 | 14.7647 |

Table A-5:  Acceptor symmetric stretch fundamental excited state energies for $K = 0$ levels of $(D_2O)_2$ (in cm$^{-1}$).

| $J$ | A1 | E1 | B1 | A2 | E2 | B2 |
|-----|-----------|-----------|-----------|-----------|-----------|-----------|
| 0 | 2668.9146 | 2668.9267 | 2668.9262 | 2670.4816 | 2670.4840 | 2670.4870 |
| 1 | 2669.2775 | 2669.2894 | 2669.2891 | 2670.8443 | 2670.8467 | 2670.8497 |
| 2 | 2670.0034 | 2670.0150 | 2670.0147 | 2671.5697 | 2671.5721 | 2671.5750 |
| 3 | 2671.0921 | 2671.1032 | 2671.1029 | 2672.6577 | 2672.6601 | 2672.6629 |
| 4 | 2672.5436 | 2672.5539 | 2672.5537 | 2674.1083 | 2674.1108 | 2674.1134 |
| 5 | 2674.3578 | 2674.3671 | 2674.3666 | 2675.9213 | 2675.9240 | 2675.9262 |
| 6 | 2676.5345 | 2676.5425 | 2676.5415 | 2678.0967 | 2678.0998 | 2678.1013 |

Table A-6.  Acceptor symmetric stretch fundamental excited state energies for $K = 1-$ levels of $(D_2O)_2$ (in cm$^{-1}$).

| $J$ | A1 | E1 | B1 | A2 | E2 | B2 |
|-----|-----------|-----------|-----------|-----------|-----------|-----------|
| 1 | 2674.2820 | 2674.2831 | 2674.2791 | 2673.7122 | 2673.7222 | 2673.7215 |
| 2 | 2675.0061 | 2675.0072 | 2675.0032 | 2674.4355 | 2674.4455 | 2674.4446 |
| 3 | 2676.0921 | 2676.0933 | 2676.0894 | 2675.5204 | 2675.5302 | 2675.5292 |
| 4 | 2677.5401 | 2677.5414 | 2677.5375 | 2676.9668 | 2676.9757 | 2676.9752 |
| 5 | 2679.3498 | 2679.3513 | 2679.3474 | 2678.7746 | 2678.7816 | 2678.7825 |

Table A-7. Acceptor symmetric stretch fundamental excited state energies
for $K$ = 1+ levels of $(D_2O)_2$ (in $cm^{-1}$).

| $J$ | A1 | E1 | B1 | A2 | E2 | B2 |
|---|---|---|---|---|---|---|
| 1 | 2674.2831 | 2674.2842 | 2674.2802 | 2673.7141 | 2673.7242 | 2673.7235 |
| 2 | 2675.0094 | 2675.0105 | 2675.0066 | 2674.4414 | 2674.4515 | 2674.4506 |
| 3 | 2676.0987 | 2676.0999 | 2676.0960 | 2675.5322 | 2675.5421 | 2675.5411 |
| 4 | 2677.5511 | 2677.5524 | 2677.5485 | 2676.9865 | 2676.9956 | 2676.9951 |
| 5 | 2679.3663 | 2679.3679 | 2679.3640 | 2678.8042 | 2678.8114 | 2678.8123 |